# Experimental and numerical analysis of tribological behavior of CrAl(Si)N films during Scratch

Zhitong Chen [a, b], Biao Feng [c, *], Yuan Xia [a], Guang Li [a, *]

[a] Institute of Mechanics, Chinese Academy of Sciences, Beijing 100190, China

[b] University of Chinese Academy of Sciences, Beijing 100049, China

[c] Department of Aerospace Engineering, Iowa State University, Ames, Iowa 50011, USA

**Abstract**

Scratch sliding tests with a $ZrO_2$ ball and CrAl(Si)N films with different Si content were conducted due to CrAlSiN films having high hardness and good wear resistance. After up to 6000 cycles the specimens were analyzed by Scanning Electron Microscopy (SEM). The friction coefficient of CrAlSiN was lower than that of CrAlN film, and the lowest friction coefficient was 0.56 at Si content of 3.7%. A corresponding three-dimensional finite element model was constructed with the help of the ABAQUS to describe the mechanical response during scratch. A comparison of experimental and computational results revealed that the small elastic deformation took place in the films and substrates; the deformation friction coefficient was negligible in comparison with the Coulomb friction coefficient; and with increasing Young's modulus, the stress concentration was more obvious in CrAlSiN than in CrAlN.

**Keywords**

Scratch, CrAlSiN, tribological behavior, finite element method

---

[*] Corresponding Author:
E–mail address: *biaofeng@iastate.edu, lghit@imech.ac.cn*

# 1. Introduction

Tribology is from the Greek word '$\tau\rho\iota\beta\varpi$' meaning 'to rub', which is the field of science and technology dealing with contacting surfaces in relative motion, which means that is related to friction, wear and lubrication [1]. It plays an important role in film fracture, wear, and material lifetime and further determines the applications. At present, the scratch sliding test is widely used to evaluate the tribological properties of film. In scratch process, the finite element method (FEM) becomes an important tool because it can describe the complex mechanical responses (e.g. stress and deformation in the entire structure) while experiments not [2]. 3-D FEM simulations of scratch tests have been successful to serve this goal [3-5], In the Coulomb friction theory the relative slippage on a contact surface initiates when the magnitude of the friction stress vector arrives at the critical value $\mu\sigma_n$, where $\sigma_n$ is the normal contact stress and $\mu$ is the Coulomb coefficient of friction (COF). The scratch COF $\mu_s$ is obtained by the ratio of tangential force $F_t$ and the normal force $F_n$. Due to deformations of the surface film, $F_t$ is not only from shear stress but from a part of critical normal stress $\sigma_n$. Likewise, a part of $F_n$ also comes from shear stress. They cause the scratch friction coefficient $\mu_s$ to be a sum of Coulomb friction coefficient $\mu$ and deformation friction coefficient $\mu_d$. With the absence of stress and strain filed, it is difficult to measure $\mu_d$, and further $\mu_s$ (dependent of $\mu_d$) is not very understandable. In some cases, when friction stress on a contact surface reaches material yield strength in shear $\tau_y$, contact slippage could take place even if the slipping condition for Coulomb friction is not satisfied [6-8]. As a result, the scratch coefficient $\mu_s$ also depends on $\tau_y$. Without the knowledge of stress field, it is unclear which factor mostly determines the values of scratch friction coefficient in experiments.

Previous experimental studies have shown that CrAlSiN films are of superhardness, excellent oxidation resistance and wear resistance [9-12]. Polcar et al. studied tribological performance of CrAlSiN coatings at high temperatures and found that adhesive problems of the coatings in the beginning of the sliding tests resulted into coating failure at the moderate temperature 600 °C. Chang et al. [13] indicated that CrAlSiN hard coatings on cemented carbide cutting tools had good wear behavior and cutting performance and Chang et al. [14] explained the influence of Si contents on tribological characteristics of CrAl(Si)N nanocomposite coatings. However, they did not describe the complex mechanical responses. In this study, we plan to point out the mechanisms of CrAl(Si)N films' mechanical responses during scratch. We try to relate the damage and geometry change of CrAl(Si)N films during scratch to the computation results of finite element model. CrAl(Si)N films with different Si content were deposited on HSS and Si wafer substrates using medium frequency pulse magnetron sputtering. The scratch sliding tests of CrAl(Si)N films with different Si content were performed on a ball-on-disc rotational wear apparatus and the ZrO2 ball was used as a counterpart material.

## 2. Experimental Procedure

The CrAl(Si)N films investigated in this research were deposited on HSS and Si wafer substrates using medium frequency pulse magnetron sputtering technique. Customized Cr and AlSi with different Si content targets in a reactive nitrogen atmosphere were used to obtain films with an increasing Si content. The HSS substrates of the disc type (30 mm in diameter and 5 mm in thickness) were cleaned in an ultrasonic cleaner using acetone and alcohol for 20 min. Mirror-polished Si and polished Cu substrate were cleaned separately in an ultrasonic cleaner using acetone and alcohol for 2 min and 20 min, respectively. The substrates were cleaned again by ion bombardment using a bias voltage of −900 V under Ar atmosphere of 1.5 Pa for 15 min.

Substrate bias voltage of -150 V and ratio $N_2$/Ar of 1:2 were used. The films were deposited from sputter sources at a working pressure of 0.5 Pa and the input power on the CrAl(Si) target was fixed at 250W. The background pressure of our apparatus was $5\times10^{-5}$ Pa.

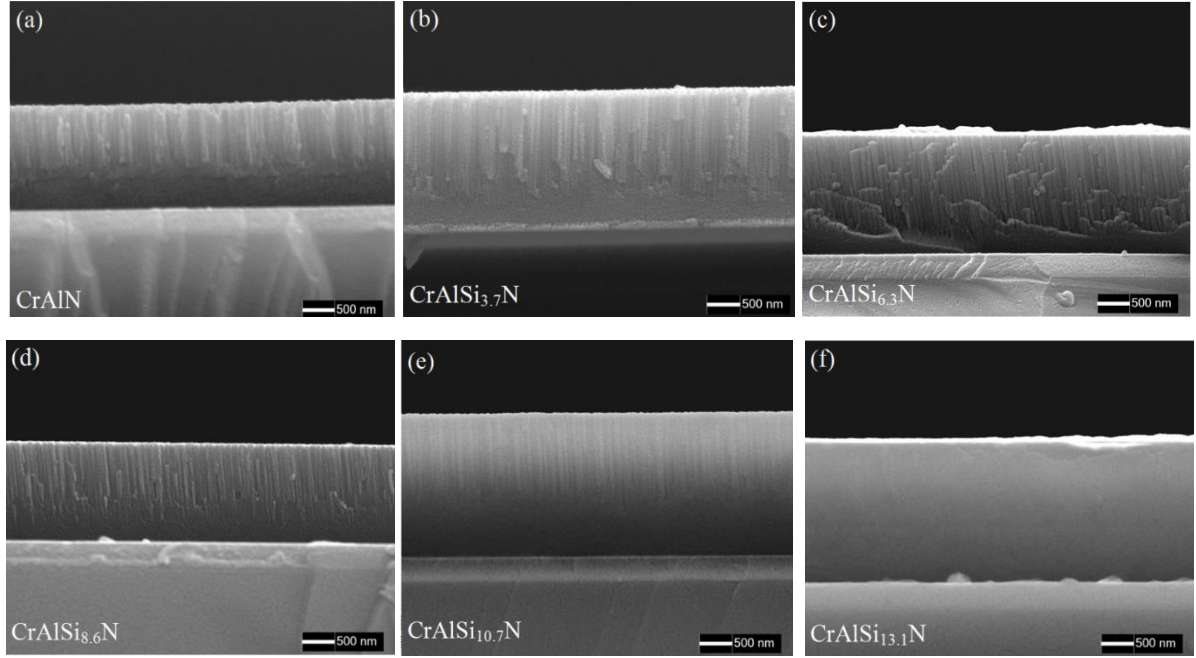

Fig. 1. Cross-sectional SEM images of CrAl(Si)N films with various Si content

Fig. 1 shows the cross-sectional micrographs of CrAl(Si)N films with different Si content that we have prepared. The columnar crystals growth direction of films is perpendicular to the film surface. The insertion of Si into nitride films may improve the densification of film and reduce the grain size of columnar films. The columnar structure of the CrAlN film switches to spherical and equiaxial grains with 13.1% Si content. In our previous studies [15], a retardation of the columnar growth by the addition of Si was verified from the elimination of (111) preferred orientation (no diffraction peaks of $SiN_x$) and the hardness of CrAlSiN films exhibited the maximum hardness value at Si content approximate 8.6% due to the microstructural change to crystal size refinement as well as solid-solution hardening.

The scratch sliding tribological tests of CrAl(Si)N films with different Si content were performed on a conventional ball-on-disc rotational wear apparatus to evaluate the friction coefficient and wear behaviors. A $ZrO_2$ ball was used as a counterpart material. The tests were conducted with a sliding speed of 0.088 m/s under a load of 1 N at ambient temperature (around 25°C) and relative humidity (25-30% RH) condition. SEM was employed to observe the morphology of the wear track after each sliding experiment.

## 3. Numerical Procedure

In the finite element model, the radius of spherical indenter was $R = 3170\ \mu m$, and thickness of film was $h = 3\ \mu m$ consistent with experiments. The thickness of the sample used in simulations is $150\ \mu m$ (in spite of 15 mm in experiments) because the results show that stresses and strains were zero in the region $(z < -100\mu m)$ and that further increasing the thickness of the substrate did not cause any effect. In the current simulations, $700\ \mu m$ was used for the length and $360\ \mu m$ was used for the width. The extra loadings were applied by following two steps: first, the indenter was compressed down to the coated surface by the normal force $F_n = 1N$; and second, a tangential displacement loading was applied to keep the indenter sliding along *x*-axis at the fixed $F_n$. The boundary conditions were as follows: 1) Coulomb friction was applied on the contact surface between the indenter and the film; 2) complete cohesion was employed on the contact surface between film and substrate; and 3) the other surfaces of the film and substrate were fixed.

In one revolution of the indenter on the films, the magnitudes of the applied tangent and normal forces were constant in the ball-on-disc wear apparatus. As a reasonable approximation, a static condition was applied and the scratch behavior was modeled by a sphere indenter scratching on a thin film coated on a thick substrate as shown in Fig. 2.

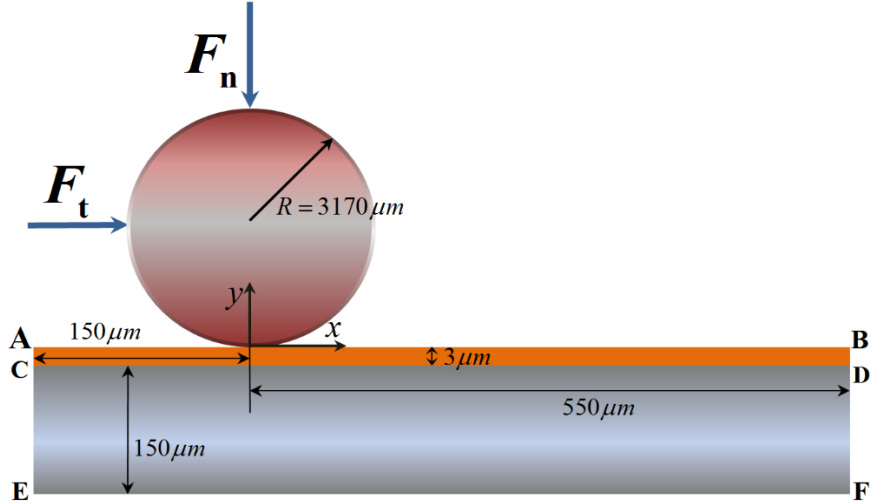

Fig. 2. Schematic representation of the scratch process on a coated substrate

As in pioneering research (e.g. in [2, 6-8, 16-18]), the indenter was also accepted as a rigid body, and the effect of this condition would be discussed. For the coating and substrate, a linearly elastic and perfect plasticity is accepted, and equations can be found in our previous publication [2]. The following material properties are used for the film: yield strength $\sigma_{y1} = 9.5 \text{GPa}$ (CrAlN) and $15.3 Gpa$ (CrAlSi$_{8.6}$N), Young's modulus $E_1 = 341 \text{GPa}$ (CrAlN) and 500Gpa (CrAlSi$_{8.6}$N); and for the HSS substrate: $\sigma_{y2} = 4.1 \text{GPa}$ and $E_2 = 200 \text{GPa}$. Poisson's ratio almost does not affect tribological behaviors (see e.g.[19]), and is taken as $v = 0.3$ for both film and substrate. Coulomb friction coefficient is taken as $\mu = 0.3$.

## 4. Results and Discussion

In the dry sliding tests, the COF for the films against ZrO$_2$ balls were measured. All specimens exhibited low values of COF at the initial running-in stage. The COF increased gradually as the sliding distances increased until stable stage was reached. The average COF was calculated over the entire sliding distance and shown in Fig. 3. The COF of CrAlN film was higher than that of the CrAlSiN films, therefore Si dissolution into CrAlN led to decreased in COF. In CrAlSiN films, the lowest COF of 0.56±0.03 for CrAlSiN film with Si addition approximate 3.7% was observed. As the Si content increased in the films, the COF gradually increased to maximum value of 0.68±0.02 at the Si content of approximate 8.6%. However, the COF of CrAlSiN films reduced with further increase in the Si content.

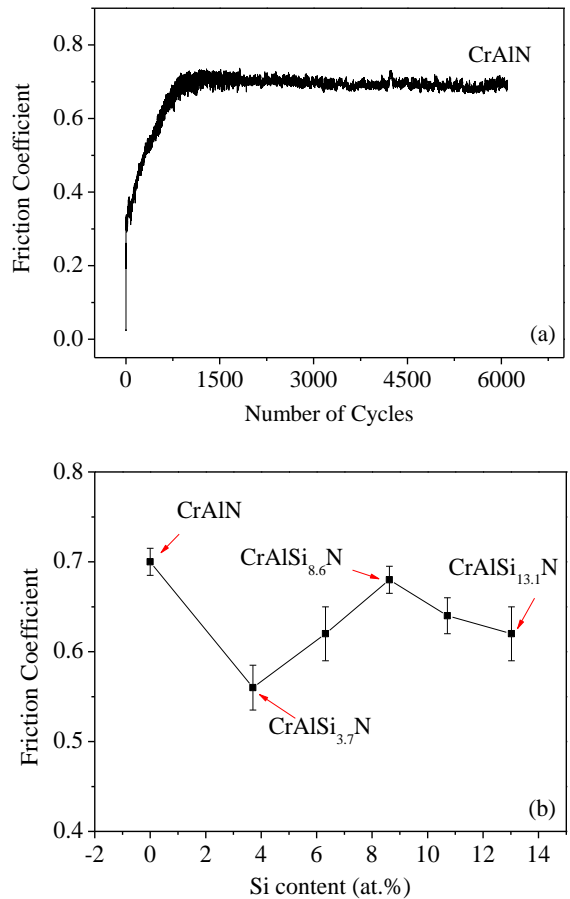

Fig. 3. Coefficient of Friction: (a) friction curves of tested CrAlN film against $ZrO_2$ ball at room temperature and (b) average COF of CrAlSiN as a function of Si content.

The film surfaces after sliding wear against $ZrO_2$ balls were examined using SEM. Wear tracks were observed on all six films against $ZrO_2$ sliding balls as shown in Fig. 4. In Fig.4 (a), there were some ploughs and debris in CrAlN film. In CrAlSiN films when Si ≤6.3%, the wear tracks of films were very smooth. When the Si content was further increased, ploughs and debris appeared and the diameter of debris increased. The CrAlSiN film was worn at the Si content of approximate 13.1%. The wear track of CrAlSi$_{3.7}$N film was much smoother than other films.

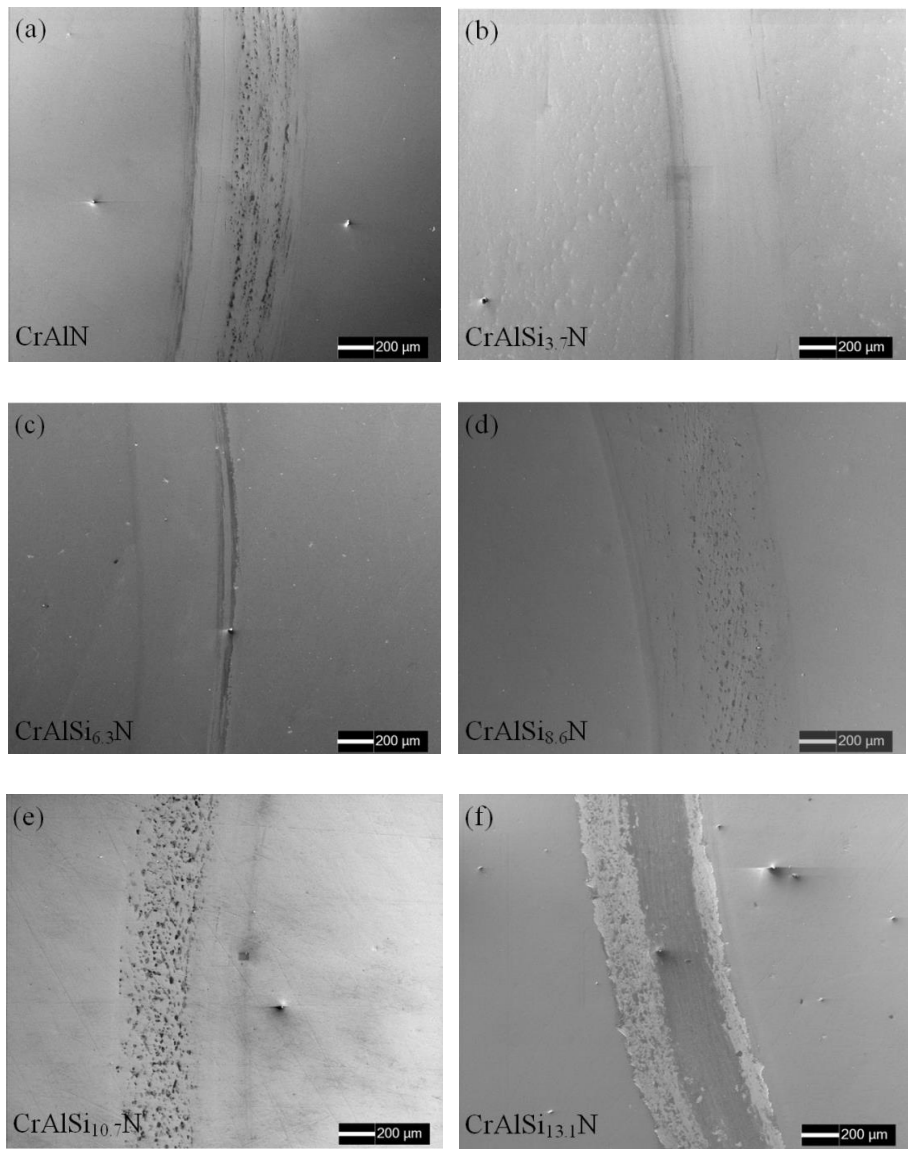

Fig. 4. The SEM wear tracks morphologies of the CrAlN and CrAlSiN films.

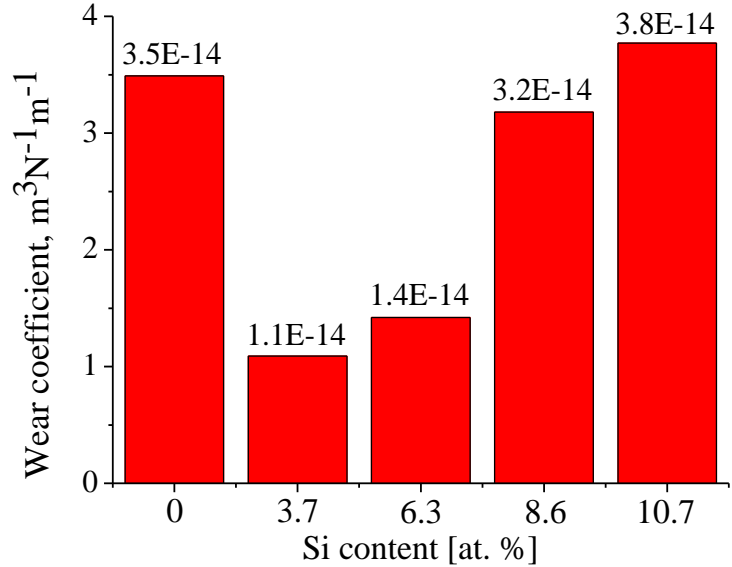

Fig. 5 Wear coefficient of CrAlSiN films deposited with different Si content

Fig. 5 shows Wear coefficient of CrAlSiN films deposited with different Si content. Wear coefficient of $CrAlSi_{13.1}N$ film was not given in Fig. 5, because the film was worn out. CrAlSiN films with low Si content ($\leq 6.3\%$) had a very low wear coefficient. Therefore, doping Si in CrAlN film could improve mechanical properties and tribological behaviors. When the Si content further increased ($\geq 8.6\%$), the wear coefficient of CrAlSiN films had a similar value with CrAlN. The high hardness of CrAlSiN films prevented significant plastic deformation under contact stress in the friction process of film so that it avoided serious ploughs. Therefore, CrAlSiN films had an excellent dry sliding and wear resistance. In CrAlSiN films, the content of amorphous $SiN_x$ phase increased with increasing Si content, while amorphous $SiN_x$ phase was much softer than the CrN phase. These results show that films can easily produce detached debris under friction force and contact stress in friction process (Fig. 3). As the Si content increased in the films, the tribological mechanism changed from abrasive wear to fatigue wear. Both CrAlN and CrAlSiN ($\geq 8.6\%$) films belonged to fatigue wear.

Due to symmetry of the plane $z=0$, which went through the center of the spherical indenter (see in Fig. 2), the results for half of the indenter-film-substrate structure with $z \leq 0$ would be presented. Since the simulation was done for half of the structure, a new boundary condition would be applied: there was no displacement along the $z$-axis $(u_3 = 0)$ on the symmetry plane $z=0$. To exclude the effect of the initial indentation, all results would be extracted at the slippage distance of the indenter 300 $\mu m$, which meant that the indenter's tip in Fig. 2 slides to the position $x=$ 300 $\mu m$. Fig. 6 represents the distribution of the effective stress $\sigma_i$ on the contact surface. For the CrAlN film the effective stress was much smaller than the material yield stress $\sigma_y$, which indicated that the material completely elastically deforms and plastic deformation did not exist. Different from Ref.[20], the residual stress was not observed in Fig. 6, because in addition to the absence of plastic deformation, elastic deformation in the current model was very small in the contact region and the region far from the contact region material deformation could automatically disappear. Without plastic deformation, plastic sliding in Ref. [6-8] also did not occur and classic elasticity laws without considering size effects by plastic strain gradient theories (see References in [21]) was reasonable for the current model. As long as it is larger than the effective stress the magnitude of yield strength does not affect the deformation at all. When the yield strength is accepted as one third of the material hardness [22], the magnitude of material hardness does not result in the difference on the scratch friction coefficient in Fig. 3 (b). Consequently, the distinctions of the stress-strain field between CrAlN and CrAlSi$_{8.6}$N were only determined by Young's modulus. With the increased in Young's modulus from CrAlN to CrAlSiN, material became harder and more difficult to deform elastically. Under the fixed normal force $F_n$=1N ($F_n$=0.5N for a half of structure), the CrAlN film was easily deformed, leading to a less stress concentration in comparison with CrAlSiN (Fig. 6). The region with large effective stress was localized at the right contact boundary in the neighborhood of point *c*, due to a large tensional stress $\sigma_{xx}$. This tensional stress was due to the large tangential friction stress at the content region If a very large normal force $F_n$ was applied to cause a severe plastic deformation, the largest effective stress would not occur in the neighborhood of point *c*, because a large $F_n$ caused a compressive stress $\sigma_{xx}$ during the extrusion of the material, which compensated the tensional stress $\sigma_{xx}$ due to the friction stress. A complete cohesion condition was used on the contact region between the film and substrate, which meant that the displacement was continuous but the effective stress was not continuous due to different Young's moduli (while normal and shear stresses are continuous).

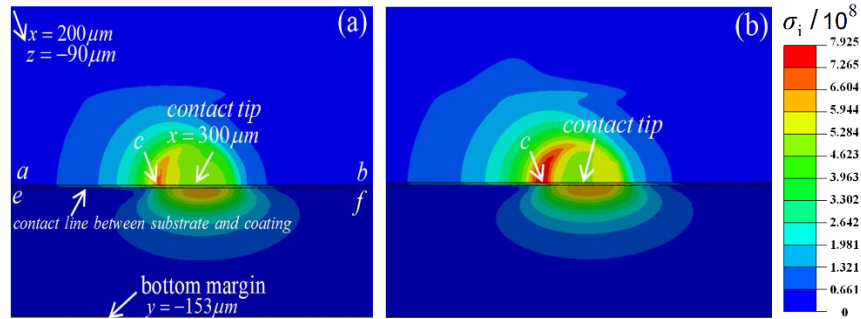

Fig. 6. The distribution of the effective stress $\sigma_i$ on the film and substrate in the current configuration for (a) CrAlN and (b) CrAlSi$_{8.6}$N with the same region. Curve line *ab* was the on *x*-axis in the undeformed configuration, and Curve *ef* was the contact line between substrate and film and with *y*=-3 $\mu m$ and *z*=0 in the undeformed configuration. Point *c* was on the curve *ab* and is the cut-off point between contact and noncontact for the indenter and film.

The distribution of the displacement along the *y*-axis $u_2$ is shown in Fig. 7 for CrAlN and CrAlSi$_{8.6}$N. With an increase in Young's modulus in the films, displacement $u_2$ did not change. As discussed before, the displacement changeed continuously from the film to the substrate, which could be seen in Fig. 9. There was no change in displacement $u_2$ along the thickness of direction in the film, which meant $u_2$ was not caused by the compressive deformation of the film but by the sinking of the substrate. Because HSS was also a "hard" material with a large Young's modulus, the displacement $u_2$ was very small in the current model. There was no displacement on the left side of the film or substrate in Fig. 7, an indication that there was no residual strain/stress after the indenter slided away. The maximum magnitude of $u_2$ in the films, $|u_2|_{max}$, was 0.1372 for CrAlN, which was slightly larger than 0.1356 for CrAlSi$_{8.6}$N, which have been caused by a larger deformation in CrAlN. Because the spherical indenter was accepted as a rigid body the contact region had the same profile as the spherical surface. Due to the similarity of the displacement $u_2$, the contact area was also very similar for CrAlN and CrAlSiN films, which could be seen in Fig. 8. In Fig. 8, the differences of distributions of normal contact pressure $\sigma_n$ between CrAlN and CrAlSiN were not very obvious and the maximum normal stress $\sigma_n$ was larger in CrAlSiN because with a smaller Young's modulus, the stress concentration was less severe. The contact radius between the indenter and film was only around 21 $\mu m$ but the radius of the indenter was 3170 $\mu m$, an indication that the contact surface was almost flat and the contribution of the deformation friction coefficient $\mu_d$ was negligible and the scratch friction coefficient $\mu_s$ was equal to the Coulomb friction coefficient $\mu$. In the current model, the order of magnitude of the deformation friction coefficient $\mu_d$ was $10^{-4}$.

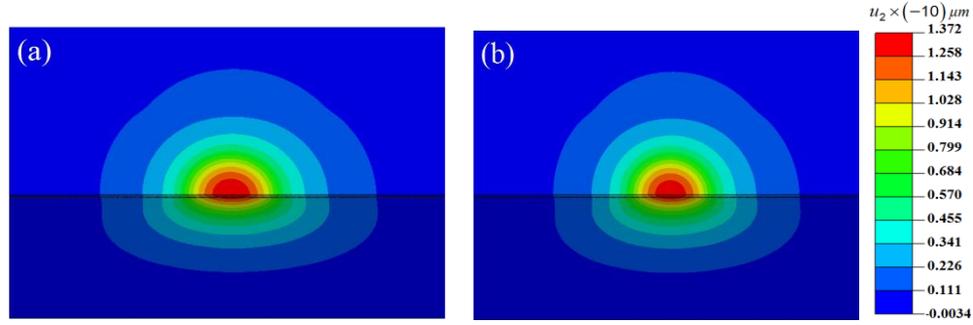

Fig. 7. the distribution of displacement of $u_2$ on the film and the substrate for the same region in Fig. (8) for (a) CrAlN and (b) CrAlSi$_{8.6}$N. The maximum magnitude of $u_2$, $|u_2|_{max}$, was 0.1372 for CrAlN and 0.1356 for CrAlSi$_{8.6}$N at the contact tip.

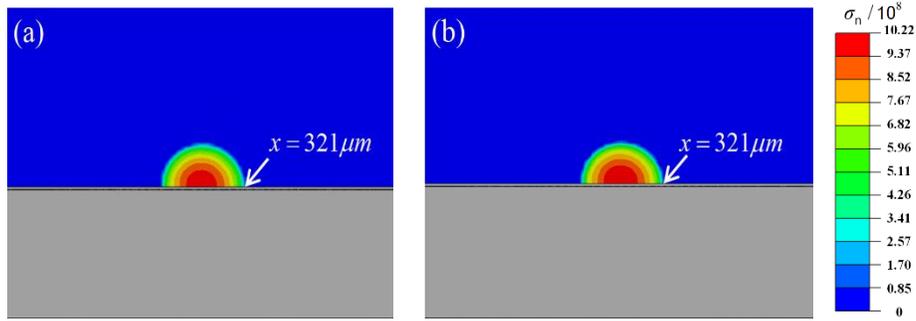

Fig. 8. the distribution of normal contact stress $\sigma_n$ on the film for (a) CrAlN and (b) CrAlSi$_{8.6}$N. The maximum of magnitude $\sigma_n$ was $1.002 \times 10^9$ GPa for CrAlN and $1.022 \times 10^9$ GPa for CrAlSi$_{8.6}$N.

While the 3D model was constructed to consider the tribological behavior, there were some challenges which could not be overcomed in the current model or in previous models [16-18, 23-25] to simulate the real contact environment. As in Ref. [23-25], the indenter was assumed to be a rigid body. When the elastic-plasticity material was utilized for the indenter ZrO$_2$, Figs. 6-8 indicated that the deformation of a sphere would also be elastic and small, because ZrO$_2$ like HSS was a very hard material. In this case, the small deformation of the indenter would increase the contact area between the indenter and film, which led to a smaller stress concentration, and further to a smaller effective stress, elastic deformations, and displacements in the film.

However, the use of the elastic-plasticity constitutive for the indenter would draw the same conclusion that the deformation of the film was completely elastic and scratch coefficient was only from the Coulomb friction coefficient. Secondly, the wear behaviors of the films were not taken into account in the models. Fig. 3(a) shows an increasing scratch friction with number of revolutions, which is due to atomic changes between contact surfaces. Initially, the contact was between $ZrO_2$ and the film CrAlN, then between the mixture of $ZrO_2$ and CrAlN, and CrAlN atoms, and finally between only CrAlN atoms. The Coulomb friction coefficient increased from $ZrO_2$-CrAlN contact to CrAlN-CrAlN contact, which led to an increasing scratch friction coefficient in Fig. 3(a) with an increasing number of circles. When the contact was only between CrAlN atoms, the Coulomb friction coefficient (equal to the scratch friction coefficient) did not change, and the curve in Fig. 3(a) became flat. Similar to hardness [17], with the increase of Si content from 0 to 8.6%, and then to 13.1%, Yong modulus initially grew and then dropped, which would cause the stress concentration to strengthen at first and then weaken. All of the process was Young modulus controlled and the yield strength did not play a role in the scratching in this paper.

## 5. Conclusions

CrAlSiN films with different Si content were deposited on HSS and Si wafer substrates using medium frequency pulse magnetron sputtering. Scratch sliding tests between a $ZrO_2$ ball and CrAl(Si)N films with different Si content were conducted. The friction coefficient of CrAlSiN was lower than that of CrAlN film and the lowest friction coefficient was 0.56 at Si content of 3.7%. The wear track of $CrAlSi_{3.7}N$ film was much smoother than other films. Also, the lowest wear rate was $1.1 \times 10^{-14} m^3 N^{-1} m^{-1}$ at Si content of 3.7%. The mechanical responses for tribological behaviors were described by a 3D finite element model was constructed using the finite element code ABAQUS. Numerical results show that the small elastic deformation took place in the films and substrates; the deformation friction coefficient was negligible in comparison with the Coulomb friction coefficient; and with increasing Young's modulus, the stress concentration was more obvious in CrAlSiN than in CrAlN.

**Acknowledgements**

The authors would like to thank the financial support of the National Nature Science Foundation of China (No: 10832011).